\documentclass[final,3p,times,numbers,sort&compress]{elsarticle}
\usepackage{graphicx}
\usepackage{subfigure}
\usepackage{amsmath}
\usepackage{booktabs}
\usepackage{color}
\usepackage{subeqnarray}

\usepackage{amssymb}
\usepackage{bm}

\usepackage[pdfpagelabels]{hyperref}

\usepackage{multirow}

\journal{}

\begin{document}

\begin{frontmatter}



\title{Some theoretical results on the second-order conservative phase field equation}

\author[bjtu]{Yang Hu\corref{cor1}}
\ead{yanghu@bjtu.edu.cn}

\author[bjtu,tsu]{Decai Li}

\cortext[cor1]{Corresponding author}

\address[bjtu]{School of Mechanical, Electronic and Control engineering, Beijing Jiaotong University, Beijing, P.R.China 100044}

\address[tsu]{State key laboratory of tribology, Tsinghua University, Beijing, P.R.China 100084}

%

\begin{abstract}

In this paper, a theoretical research on the second-order conservative phase field (SOCPF) equation is presented. The theoretical results include the following three aspects. First, three new derivation methods for the SOCPF equation are given. The SOCPF equation can be viewed as the gradient flow, the special diffusion equation and the diffuse interface form of a sharp interface formulation for the piecewise constant function, respectively. These derivation methods help us to understand the SOCPF equation at different perspectives. Second, the conservation's properties of the solution of SOCPF equation are studied. Compared with the Cahn-Hilliard equation and the Allen-Cahn equation, it is found that the solution of SOCPF equation satisfies more conservation laws. Third, the wetting boundary condition for the SOCPF equation is investigated. We find that the no-flux boundary condition is equivalent to the wetting boundary condition for two-component phase field model. Moreover, applying the no-flux boundary conditions for $N$-component phase field model, we give a set of wetting boundary conditions for $N$ phase field parameters.

\end{abstract}


\end{frontmatter}


\section{Introduction}
\label{}

Immiscible multiphase flows can be frequently found in many engineering and science fields, ranging from petroleum industry, heat exchanger, aerospace industry, chemical process and micro electric mechanical system. How to effectively capture the time-evolving interface and accurately calculate the surface tension makes the modeling of such flows a rather challenging work. Up to now, people have developed many numerical methods to deal with the multiphase flows, including the volume of fluid method \cite{Hirt:1981}, the level set method \cite{Osher:1988,Sussman:1994}, the front tracking method \cite{Tryggvason:2001,Unverdi:1992} and the phase field method \cite{Anderson:1998,Yue:2004,Ding:2007}. The phase field method, due to its solid physical background and simple calculation process, has received more and more attention in recent years.

In the phase field method, the phase interface is replaced by a thin but nonzero-thickness transitional region. A scalar function, the order parameter, is used to identify the different phases. Three types of equations are usually applied to capture the moving interfaces in the phase field method: the Cahn-Hilliard (C-H) equation \cite{Cahn:1958}, the Allen-Cahn (A-C) equation \cite{Allen:1976} and the SOCPF equation\cite{Sun:2007,Chiu:2011}. The difference between three equations mainly lies in the different diffusion terms. In the C-H equation, the diffusive flow rate is taken to be proportional to the gradient of the chemical potential\cite{Jacqmin:1999}. The phase field model based on the C-H equation has been used for phase separation \cite{Badalassi:2003}, Hele-Shaw flows \cite{Folch:2005}, vesicle dynamics \cite{Biben:2005}, moving contact lines \cite{Seppecher:1996,Jacqmin:2000}, two-phase flows with soluble surfactants \cite{Teigen:2011} and two-phase ferrofluid flows \cite{Nochetto:2016}. Because the C-H equation is a fourth order partial differential equation, it is difficult to develop high accuracy and efficient numerical method. Although the C-H equation can conserve the total mass, it can not preserve the volume of each phase \cite{Yue:2007}. To avoid the treatment of the high order derivatives, the A-C equation which only requires second-order derivatives is also used to track the interface. The original A-C equation can not be written as a conservative form. So a Lagrange multiplier should be introduced to ensure the mass conservation law \cite{Bronsard:1997,Brassel:2011}. Moreover, the A-C equation provides an approximation to the motion by the volume-preserving mean curvature flow. It denotes that the volume-preserving property is satisfied. However, the local curvature of the phase interface tends to the mean curvature in the A-C equation. As a result, it may fail to capture some small feature of the flow field \cite{Hu:2019pre}.

In addition to the C-H equation and the A-C equation, the SOCPF equation has received more and more attention in recent years. The SOCPF equation was first proposed by Sun and Beckmann\cite{Sun:2007}. It was derived from an interface advection equation by expressing the interface normal and curvature in terms of a hyperbolic tangent phase-field profile across the interface \cite{Sun:2007}. Like the C-H equation, the SOCPF equation can be reformulated into the conservative form\cite{Chiu:2011}. Like the A-C equation, the SOCPF equation is a second-order equation and it is simpler than the C-H equation. The SOCPF equation has been used in many numerical simulations \cite{Geier:2015,Hu:2018pre,Hu:2019ijhmt,Chai:2018,Liang:2018,He:2019}. These numerical practices demonstrated its advantages against the two other phase field models.

In this paper, some theoretical results for the SOCPF equation are presented. The three aspects of contributions of this paper are: first, we give three new derivations for the SOCPF equation. Such equation can be viewed as the motion by a gradient flow, a special diffusion equation, or the diffuse interface form of a sharp interface formulation for piecewise constant function. We can see this equation from different perspectives. Second, the conservation property of the solution of the SOCPF equation is investigated. It is found that the conservation laws of the total mass, the volume of each phase and the surface-area of phase interface are satisfied. Third, the relationship between the mass conservation law and the wetting boundary condition is studied. Different from the C-H equation, it is found the mass conservation law and the wetting boundary condition for the SOCPF equation are tied-up. We prove that the no-flux boundary condition is equivalent to the wetting boundary condition in two-phase phase field model. Furthermore, we also consider the wetting boundary conditions for N-component phase field model. Utilizing the no-flux boundary conditions, we derive a set of wetting boundary conditions for N phase field parameters.

\section{Derivations of the SOCPF equation}

The SOCPF equation is written as
\begin{eqnarray}\label{SOCPFEQ}
   \frac{\partial \phi}{\partial t}=M \nabla \cdot[\nabla \phi-\frac{4\phi(1-\phi)}{\xi}\textbf{n}]
\end{eqnarray}
where $\phi$ is the phase field parameter with respect to the position $\textbf{x}$ and time $t$. $\xi$ is a parameter related to the interface thickness. $M$ is the mobility. $\textbf{n}=\frac{\nabla \phi}{|\nabla \phi|}$ is the unit vector normal to the interface. At the equilibrium state, the order parameter along the normal direction has the following form
\begin{eqnarray}\label{normalProfile}
    \phi(z)=0.5+0.5\tanh(\frac{2z}{\xi}),
\end{eqnarray}
where $z$ is the signed distant to the interface.

First of all, we introduce the existing derivation of the SOCPF equation \cite{Sun:2007,Chiu:2011}. Consider the curvature-driven level set equation
\begin{eqnarray}\label{curvature}
   \frac{\partial \phi}{\partial t}=MK|\nabla \phi|,
\end{eqnarray}
where $K=\nabla \cdot \textbf{n}$ is the curvature. The term on the right hand side of the above equation can be calculated as
\begin{eqnarray}
   K|\nabla \phi|=\nabla \cdot (\frac{\nabla \phi}{|\nabla \phi|})|\nabla \phi|=\frac{\nabla^2 \phi |\nabla \phi|-\nabla |\nabla \phi|\cdot \nabla \phi}{|\nabla \phi|}=\nabla^2 \phi-\nabla |\nabla \phi|\cdot \textbf{n}.
\end{eqnarray}
Then the Eq. (\ref{curvature}) can be written as
\begin{eqnarray}
   \frac{\partial \phi}{\partial t}=M(\nabla^2 \phi-\nabla |\nabla \phi|\cdot \textbf{n}).
\end{eqnarray}
For multiphase flows problem, there is no curvature-driven interface motion. To this end, the right hand side of above equation should minus the term $MK|\nabla \phi|$, i.e. $M|\nabla \phi|\nabla \cdot \textbf{n}$
\begin{eqnarray}
   \frac{\partial \phi}{\partial t}=M[\nabla^2 \phi-\nabla (|\nabla \phi|)\cdot \textbf{n}-|\nabla \phi|\nabla \cdot \textbf{n}]=M\nabla \cdot[\nabla \phi-|\nabla \phi|\textbf{n}].
\end{eqnarray}
By using the property of sigmoid function $|\nabla \phi|=\frac{4\phi(1-\phi)}{\xi}$, the Eq. (\ref{SOCPFEQ}) is obtained.

In addition to the above derivation method provided by Chiu and Lin \cite{Chiu:2011}, we also give three kinds of alternative methods to derive the SOCPF equation. It will help deepen understanding of this equation.

\subsection{The gradient flow method}

Both the C-H equation and the A-C equation can be interpreted as the gradient flow \cite{Shen:2011}. The energy functional is
\begin{eqnarray}
    W_p(\phi,\nabla \phi)=\int_\Omega [\frac{1}{2}|\nabla \phi|^2+\frac{1}{4\xi^2}\phi^2(\phi-1)^2]d\textbf{x},
\end{eqnarray}
The gradient flow equation is
\begin{eqnarray}
    \frac{\partial \phi}{\partial t}=-M\frac{\delta W_p}{\delta \phi}.
\end{eqnarray}
Calculating the variational term $\frac{\delta W_p}{\delta \phi}$ in $H^{-1}$ space, the C-H equation is achieved
\begin{eqnarray}
    \frac{\partial \phi}{\partial t}=M\nabla^2 \mu_\phi,
\end{eqnarray}
where $\mu_\phi$ is the chemical potential
\begin{eqnarray}
    \mu_\phi=\frac{8}{\xi^2}\phi(\phi-1)(\phi-0.5)-\nabla^2 \phi.
\end{eqnarray}
Calculating the variational term $\frac{\delta W_p}{\delta \phi}$ in $L^2$ space, the A-C equation is obtained
\begin{eqnarray}
    \frac{\partial \phi}{\partial t}=M[\nabla^2 \phi-\frac{8}{\xi^2}\phi(\phi-1)(\phi-0.5)],
\end{eqnarray}

To derive the SOCPF equation in gradient flow framework, we define the following quadratic functional
\begin{eqnarray}
    W_g(\phi,\nabla \phi)=\int_\Omega \frac{1}{2}[|\nabla \phi|-\frac{4\phi(1-\phi)}{\xi}]^2d\textbf{x}.
\end{eqnarray}
Obviously, minimizing the functional $W_g$, the relation $|\nabla \phi|=\frac{4\phi(1-\phi)}{\xi}$ will be enforced.

Similarly, consider the following gradient flow equation
\begin{eqnarray}\label{gradflow}
    \frac{\partial \phi}{\partial t}=-M\frac{\delta W_g}{\delta \phi}.
\end{eqnarray}
If we take the variational derivative $\frac{\delta W_g}{\delta \phi}$ in $L^2$ space, one can obtain
\begin{eqnarray}\label{OriEq}
    \frac{\partial \phi}{\partial t}=M\nabla \cdot \{[|\nabla \phi|-\frac{4\phi(1-\phi)}{\xi}]\frac{\nabla \phi}{|\nabla \phi|}\}+M[|\nabla \phi|-\frac{4\phi(1-\phi)}{\xi}]\frac{4(1-2\phi)}{\xi}.
\end{eqnarray}
The existence of the second term on the right hand side of the above equation makes it can not be written as a conservative form. In other word, we can not define a flux $\textbf{J}$ to express the right hand side of the above equation as $\nabla \cdot \textbf{J}$. Fortunately, $|\nabla \phi|$ is approximately equal to $\frac{4\phi(1-\phi)}{\xi}$ in the phase field simulation. As a result, the second term on the right hand side of the above equation can be neglected. Then the above equation can be written as the following conservative form
\begin{eqnarray}\label{PreEq}
    \frac{\partial \phi}{\partial t}=M\nabla \cdot \{[|\nabla \phi|-\frac{4\phi(1-\phi)}{\xi}]\frac{\nabla \phi}{|\nabla \phi|}\}.
\end{eqnarray}
The SOCPF equation is obtained. 

\subsection{The diffusion equation with special diffusion term}

The SOCPF equation can be considered as a special diffusion equation. As stated above, the profile of the phase field parameter along the normal direction of interface is predefined as the Eq. (\ref{normalProfile}). From the distribution of the profile, we can know that the diffusion coefficient along the normal direction should be zero. In fact, if the diffusion coefficient along the normal direction is not equal to zero, the diffusion along the normal direction will occur. It will lead to the profile of $\phi$ deviates from the Eq. (\ref{normalProfile}). So we can conclude that there is no flux along the normal direction. It indicates that the flux vector of $\phi$ should be the standard flux vector $M\nabla \phi$ minus the flux vector in the normal direction $M\textbf{n}\frac{\partial\phi}{\partial n}$, i.e. $M(\nabla \phi-\textbf{n}\frac{\partial\phi}{\partial n})$. As a result, the governing equation of $\phi$ should be written as
\begin{eqnarray}\label{TangentialEq}
   \frac{\partial \phi}{\partial t}=M \nabla \cdot(\nabla \phi-\textbf{n}\frac{\partial\phi}{\partial n})=M \nabla \cdot[(\textbf{I}-\textbf{n}\textbf{n})\cdot\nabla \phi],
\end{eqnarray}
where $\textbf{I}$ is the unit tensor. The anisotropic diffusion coefficient matrix is $M(\textbf{I}-\textbf{n}\textbf{n})$. Its eigenvalues are $M, 0$ for two-dimensional case and $M, M, 0$ for three-dimensional case, respectively. Moreover, the normal gradient of $\phi$ satisfies
 \begin{eqnarray}\label{normalGrad}
   \frac{\partial \phi}{\partial n}=|\textbf{n}\cdot \nabla \phi|=\frac{4\phi(1-\phi)}{\xi}.
\end{eqnarray}
Substituting Eq. (\ref{normalGrad}) into Eq. (\ref{TangentialEq}), the phase field equation is obtained.

In addition to the anisotropic diffusion, the diffusion term in the phase field equation can be also interpreted as an special isotropic diffusion. It is pointed out that the phase field equation can be also rewritten as
\begin{eqnarray}
   \frac{\partial \phi}{\partial t}=\nabla \cdot(D_e \nabla \phi),
\end{eqnarray}
where $D_e$ is the effective diffusion coefficient
\begin{eqnarray}
   D_e=1-\frac{4\phi(1-\phi)}{\xi|\nabla \phi|}.
\end{eqnarray}
If $|\nabla\phi|>\frac{4}{\xi}\phi(1-\phi)$, we can obtain $D_e>0$. The diffusion process is a down-hill diffusion. This diffusion effect makes $\phi$
more even and the gradient $|\nabla \phi|$ will be reduced. On the other hand, $|\nabla\phi|<\frac{4}{\xi}\phi(1-\phi)$ leads to $D_e<0$. The diffusion process is a up-hill diffusion. It makes $|\nabla \phi|$ increasing. So $|\nabla \phi|=\frac{4}{\xi}\phi(1-\phi)$ is enforced.

\subsection{Diffuse interface form of a sharp interface formulation for the piecewise constant function}

In this subsection, we establish a sharp interface formulation for the piecewise constant function and give its diffuse interface form. We prove that the diffuse interface form is the SOCPF equation. Consider a domain $\Omega$ which is divided into the two subdomains $\Omega^+$ and $ \Omega^-$. We define two scalar constant function $\psi^+(\textbf{x},t)$ and $\psi^-(\textbf{x},t)$ as
\begin{eqnarray}
    && \psi^+(\textbf{x},t)=1, \textbf{x} \in \Omega^+,\\
    && \psi^-(\textbf{x},t)=0, \textbf{x} \in \Omega^-.
\end{eqnarray}
Obviously, $\psi^i (i=+,-)$ satisfy the following diffusion equation
\begin{eqnarray}\label{twodiffusion}
    \frac{\partial \psi^i}{\partial t}=M \nabla^2 \psi^i.
\end{eqnarray}
To get the sharp interface formulation, we define a scalar function $\psi$ in whole domain $\Omega$
\begin{eqnarray}
    \psi(\textbf{x},t)=\left\{
                         \begin{array}{ll}
                           \psi^+(\textbf{x},t), &  \textbf{x} \in \Omega^+, \\
                           \psi^-(\textbf{x},t), &  \textbf{x} \in \Omega^-.
                         \end{array}
                       \right.
\end{eqnarray}
The jump conditions of $\psi$ and its normal derivative on the interface $\Gamma$ satisfy
\begin{eqnarray}
    &&\psi^+=\psi^-+1,\label{Jump1}\\
    &&\frac{\partial\psi^+}{\partial n}=\frac{\partial\psi^-}{\partial n}.\label{Jump2}
\end{eqnarray}

By establishing the weak form of the proposed equation and choosing the special test function, we can prove the two diffusion equations (\ref{twodiffusion}) with two interface jump conditions (\ref{Jump1}) and (\ref{Jump2}) are equivalent to the following sharp interface formulation
\begin{eqnarray}\label{SharpEq}
    \frac{\partial \psi}{\partial t}=M \nabla^2 \psi-M \nabla \cdot\int_\Gamma \textbf{n}\delta(\textbf{x}-\textbf{X})dS,
\end{eqnarray}
where $\textbf{X}$ is the coordinate of point on the interface $\Gamma$. Note that $\psi$ actually is a Heaviside function. We replace $\psi$ and $\delta$ by the discrete forms $\psi_\xi$ and $\delta_\xi$, respectively. The diffuse interface formulation is
\begin{eqnarray}
    \frac{\partial \psi_\xi}{\partial t}=M \nabla^2 \psi_\xi-M \nabla \cdot (\textbf{n}\delta_\xi).
\end{eqnarray}
Especially, when we choose the following forms
\begin{eqnarray}
    &&\phi=\psi_\xi(z)=0.5+0.5\tanh(\frac{2z}{\xi}),\\
    &&\delta_\xi(z)=\frac{d \psi_\xi(z)}{dz}=\frac{4\psi_\xi(1-\psi_\xi)}{\xi},
\end{eqnarray}
we obtain the SOCPF equation again.

\section{The conservative laws for the SOCPF equation}

In this section, we discuss some conservation's properties of the solution of the SOCPF equation. In the phase field model, the conservation of the total mass and the volume of each phase is of particular concern. Although the multi-component fluids are assumed to be incompressible, note that the total mass conservation is one thing and the volume preservation of each phase is another thing. For example, when the no-flux boundary condition is applied, the C-H equation can conserve the total mass. Yue et al. presented a detailed theoretical analysis for the C-H equation \cite{Yue:2007}. They found that the bulk phase concentrations tend to shift from equilibrium concentration values. As a result, the radius of the drop will decrease even go into zero. The volume loss is proportional to the Cahn number $Cn = \xi/L$, where $L$ is the reference length. For the SOCPF equation, the equilibrium profile of the phase field parameter is enforced. Compared with the C-H equation, more conservation laws for the SOCPF equation can be achieved.

Like the C-H equation, the SOCPF equation can be written as the conservative form. In fact, we define a flux $\textbf{J}$
\begin{eqnarray}
   \textbf{J}=\nabla \phi-\frac{4\phi(1-\phi)}{\xi}\frac{\nabla \phi}{|\nabla \phi|}.
\end{eqnarray}
Then the SOCPF equation can be written as
\begin{eqnarray}
   \frac{\partial \phi}{\partial t}=M \nabla \cdot \textbf{J}.
\end{eqnarray}
When the no-flux boundary condition is utilized,i.e.
\begin{eqnarray}
   \textbf{J} \cdot \textbf{n}_w=0,
\end{eqnarray}
we can get
\begin{eqnarray}
   \frac{d}{dt}\int_\Omega \phi d\textbf{x}=\int_\Omega \frac{\partial \phi}{\partial t} d\textbf{x}=\int_\Omega \nabla \cdot \textbf{J} d\textbf{x}=\int_{\partial\Omega} \textbf{J}\cdot\textbf{n}_w dS=0.
\end{eqnarray}
The mass conservation law holds.

Then we consider the volume conservation and the surface-area conservation. The Reynolds transport theorem for any scalar $c$ is
\begin{eqnarray}
    \frac{d}{dt}\int_{\Omega_1} c d\textbf{x}=\int_{\Omega_1} \frac{\partial c}{\partial t} d\textbf{x}+\int_\Gamma c V_ndS,
\end{eqnarray}
where domain $\Omega_1$ is enclosed by the interface $\Gamma$. $V_n=\textbf{V}\cdot \textbf{n}$ is the component of self-generated velocity of interface motion in the normal direction. Let $c=1$, we can obtain the time derivative of the volume $Vol(\Omega_1)$
\begin{eqnarray}
    \frac{d}{dt}Vol(\Omega_1)=\frac{d}{dt}\int_{\Omega_1} d\textbf{x}=\int_\Gamma V_ndS.
\end{eqnarray}
Moreover, for the material volume element $dSdn$, we have
\begin{eqnarray}
    \frac{\partial(dSdn)}{\partial t}=\nabla \cdot \textbf{V} dSdn.
\end{eqnarray}
The time derivative of the material surface element $dS$ is given by
\begin{eqnarray}
    &&\frac{\partial(dS)}{\partial t}=\nabla \cdot \textbf{V} dS-\frac{\partial (\textbf{V}\cdot \textbf{n})}{\partial n}dS \nonumber\\
    &&=\nabla_s \cdot \textbf{V}dS=(\nabla_s \cdot \textbf{V}_s+\nabla_s \cdot \textbf{V}_n)dS\nonumber\\
    &&=(\nabla_s \cdot \textbf{V}_s+ V_n\nabla_s \cdot \textbf{n}+\textbf{n} \cdot \nabla_s V_n)dS=(\nabla_s \cdot \textbf{V}_s+ K V_n)dS,
\end{eqnarray}
where $\nabla_s=\nabla-(\textbf{n}\cdot \nabla)\textbf{n}$ is the surface gradient operator. $\textbf{V}_s$ is the tangential component and normal component of the vector $\textbf{V}$. As a result, the time derivative of the surface-area $A(\Gamma)$ is calculated as
\begin{eqnarray}
    \frac{d}{dt}A(\Gamma)=\frac{d}{dt}\int_\Gamma dS=\int_\Gamma (\nabla_s \cdot \textbf{V}_s+ K V_n)dS=\int_\Gamma K V_n dS.
\end{eqnarray}

The interface tracking equation with the seft-generated velocity of interface motion can be written as
\begin{eqnarray}
    \frac{\partial \phi}{\partial t}=-V_n|\nabla \phi|.
\end{eqnarray}
The SOCPF equation can be expressed as the above form, where $V_n$ is defined as
\begin{eqnarray}
    V_n=-\frac{M}{|\nabla \phi|}\nabla \cdot [\nabla \phi-\frac{4\phi(1-\phi)}{\xi}\frac{\nabla \phi}{|\nabla \phi|}].
\end{eqnarray}
Because $|\nabla \phi|=\frac{4\phi(1-\phi)}{\xi}$ is enforced, $V_n$ is basically zero, i.e.
\begin{eqnarray}
    V_n=-\frac{M}{|\nabla \phi|}\nabla \cdot [(1-\frac{4\phi(1-\phi)}{|\nabla \phi|\xi})\nabla \phi]=0.
\end{eqnarray}
Then the two conservation laws are obtained
\begin{eqnarray}
    &&\frac{d}{dt}Vol(\Omega_1)=0,\\
    &&\frac{d}{dt}A(\Gamma)=0.
\end{eqnarray}
Compared with the C-H equation and the A-C equation, it can be found that the solution of SOCPF equation has more conservation properties.

\section{The wetting boundary condition for the SOCPF model}

In this section, we discuss the relationship between the mass conservation law and the wetting boundary condition for the SOCPF model. As we know, for the C-H equation, the following two boundary conditions are usually used in the moving contact line problem \cite{Lee:2010}
\begin{eqnarray}
    &&\nabla \mu_\phi \cdot \textbf{n}_w=0,\\
    &&\nabla \phi \cdot \textbf{n}_w=\frac{4\phi(1-\phi)}{\xi}\cos(\theta^{eq}), \label{twophasewetting}
\end{eqnarray}
where $\theta^{eq}$ is the equilibrium contact angle. As stated above, the first equation is the no-flux boundary condition which related to the mass conservation. The second equation is the wetting boundary condition which related to the force balance on the moving contact line. It can be seen the two equations are mutually independent. In fact, the C-H equation is a fourth-order equation. It needs two boundary conditions to ensure the well-posedness of the equation. So the mass conservation law and the wetting boundary condition have no relation to one another in the C-H phase field model. Different from the C-H equation, the SOCPF is a second-order equation. To solve it, only one boundary condition is needed. So only one of the no-flux condition and the wetting boundary condition can be applied on the outer boundary. It is necessary to investigate the compatibility between the two boundary conditions. In fact, when the no-flux boundary condition is utilized, we have
\begin{eqnarray}
   \textbf{J} \cdot \textbf{n}_w=[\nabla \phi-\frac{4\phi(1-\phi)}{\xi}\textbf{n}] \cdot \textbf{n}_w=0.
\end{eqnarray}
Based on the geometric relationship $\textbf{n} \cdot \textbf{n}_w=\cos(\theta^{eq})$, one can obtain the wetting boundary condition (\ref{twophasewetting}). It denotes that the mass conservation law and the wetting boundary condition can be satisfied at the same time. We can conclude that the mass conservation law and the wetting boundary condition are tied-up for the SOCPF model.

Note that only the two-component phase field model is discussed above. Based on the relationship between the no-flux boundary condition and the wetting boundary condition, we will give the wetting boundary conditions for phase field model for N phase immiscible fluids.  For the two-phase flow model, only a phase field parameter $\phi$ is used. For the N-component phase field model, N phase field parameters $\phi_i(i=1,2,\cdots,N)$ should be used. To ensure the volume constraint, we have
\begin{eqnarray}
    \sum_i^{N} \phi_i=1.
\end{eqnarray}
Each phase field parameter $\phi_i$ is also governed by a convection-diffusion equation with a Lagrange multiplier $\beta$
\begin{eqnarray}\label{NphaseEq}
   \frac{\partial \phi_i}{\partial t}=M \nabla \cdot[\nabla \phi_i-\frac{4\phi_i(1-\phi_i)}{\xi}\frac{\nabla \phi_i}{|\nabla \phi_i|}]+M\beta.
\end{eqnarray}
Summing the Eq. (\ref{NphaseEq}) over $i$ from $1$ to $N$ leads to
\begin{eqnarray}
   \beta=\frac{1}{N}\nabla \cdot[\sum_{j=1}^N\frac{4\phi_j(1-\phi_j)}{\xi}\frac{\nabla \phi_j}{|\nabla \phi_j|}].
\end{eqnarray}
So we can also define the flux $\textbf{J}_i$
\begin{eqnarray}
    \textbf{J}_i=\nabla \phi_i-\frac{4\phi_i(1-\phi_i)}{\xi}\frac{\nabla \phi_i}{|\nabla \phi_i|}+\frac{1}{N}\sum_{j=1}^N\frac{4\phi_j(1-\phi_j)}{\xi}\frac{\nabla \phi_j}{|\nabla \phi_j|}.
\end{eqnarray}
Then the Eq. (\ref{NphaseEq}) can be written as
\begin{eqnarray}
   \frac{\partial \phi_i}{\partial t}=M \nabla \cdot \textbf{J}_i.
\end{eqnarray}
For the N-phase moving contact line problem, the interfacial wettability can be expressed in terms of static contact angles $\theta_{ij}^{eq}(i\neq j)$, where $\theta_{ij}^{eq}$ is the angle between fluid interface $\Gamma_{ij}$ and the solid interface. Moreover, it is easy to know $\theta_{ij}^{eq}+\theta_{ji}^{eq}=\pi$.

Then we try to derive a wetting boundary condition for $\phi_i$. When the no-flux boundary condition for $\phi_i$ is used, i.e.
\begin{eqnarray}
   \textbf{J}_i \cdot \textbf{n}_w=0,
\end{eqnarray}
we can obtain
\begin{eqnarray}
   &&\nabla \phi_i\cdot \textbf{n}_w=\frac{\partial\phi_i}{\partial n_w}=\frac{4\phi_i(1-\phi_i)}{\xi}\textbf{n}_i \cdot \textbf{n}_w-\sum_{j=1}^N\frac{4\phi_j(1-\phi_j)}{\xi}\textbf{n}_j \cdot \textbf{n}_w.
\end{eqnarray}
The key issue is how to calculate the vector inner product $\textbf{n}_i \cdot \textbf{n}_w$. At the intersection between the interface $\Gamma_{ij}$ and solid interface, $\textbf{n}_i \cdot \textbf{n}_w$ should be $\cos(\theta_{ij}^{eq})$. Motivated by Zhang et al. \cite{Zhang:2016}, $\textbf{n}_i \cdot \textbf{n}_w$ can be calculated by a linear interpolation method
\begin{eqnarray}
   \textbf{n}_i \cdot \textbf{n}_w=\sum_{j\neq i}^N\frac{\phi_j}{1-\phi_i}\cos(\theta_{ij}^{eq}).
\end{eqnarray}
As a result, we have
\begin{eqnarray}\label{Nphasewetting}
   &&\nabla \phi_i\cdot \textbf{n}_w=\frac{\partial\phi_i}{\partial n_w} \nonumber\\
   &&=\frac{4\phi_i}{\xi}\sum_{j\neq i}^N\phi_j\cos(\theta_{ij}^{eq})-\sum_{j=1}^N\frac{4\phi_j(1-\phi_j)}{\xi}\sum_{k\neq j}^N\frac{\phi_k}{1-\phi_j}\cos(\theta_{jk}^{eq})\nonumber\\
   &&=\frac{4\phi_i}{\xi}\sum_{j\neq i}^N\phi_j\cos(\theta_{ij}^{eq}).
\end{eqnarray}
The above equations can be used to treat the wetting boundary conditions in moving contact line problem with N-component immiscible fluids. At the intersection between the interface $\Gamma_{ij}$ and solid interface, we have $\phi_i+\phi_j=1$ and $\phi_k=0 (k\neq i, k\neq j)$. The above equation reduces to
\begin{eqnarray}
    \nabla \phi_i\cdot \textbf{n}_w=\frac{4}{\xi}\phi_i\phi_j\cos(\theta_{ij}^{eq})=\frac{4}{\xi}\phi_i(1-\phi_i)\cos(\theta_{ij}^{eq}).
\end{eqnarray}
Obviously, the two-phase wetting boundary condition is obtained. It denotes the proposed boundary conditions satisfy the reduction-consistency principle.

\section{Conclusion}
    In this paper,a theoretical study of the SOCPF equation is performed. We show that this equation can be derived from three different methods. By introducing a quadratic functional, the SOCPF equation can be treated as a gradient flow. Then we prove that the SOCPF equation can be also viewed as the surface diffusion flow. Furthermore, we give a sharp interface formulation for the piecewise constant function. The SOCPF equation can be obtained from the diffuse interface form of the sharp interface formulation. The conservation property of the solution of the SOCPF equation is also studied. It is found that It is found that the conservation laws of the total mass, the volume of each phase and the surface-area of phase interface are satisfied. Finally, the relationship between the mass conservation law and the wetting boundary condition for the SOCPF equation is investigated. It is found that the no-flux boundary condition is equivalent to the wetting boundary condition. Based on this relationship, we give a set of wetting boundary conditions for the N-component SOCPF equations.

\section*{Acknowledgments}
 \label{}

This work is supported by the National Natural Science Foundation of China (Grant No. 11802159).

\section*{Appendix: Derivation of the sharp interface formulation for piecewise constant function}
\setcounter{equation}{0}
\renewcommand\theequation{A.\arabic{equation}}

In this section, we give the proof of equivalence of the two diffusion equations (\ref{twodiffusion}) with two interface jump conditions (\ref{Jump1}) and (\ref{Jump2}) the sharp interface formulation (\ref{SharpEq}). When $\textbf{x}\neq \textbf{X}$, we have $\delta(\textbf{x}-\textbf{X})=0$. The sharp interface formulation reduces to the simple diffusion equation (\ref{twodiffusion}). To obtain the jump conditions (\ref{Jump1}) and (\ref{Jump2}), the weak form of the sharp interface formulation is applied. Let us take a banded domain $\Omega_\epsilon$ enclosing the interface $\Gamma$ with the outer and inner domain $\partial\Omega_\epsilon^+$ and $\partial\Omega_\epsilon^-$. Here $\epsilon$ is a small distance from $\Gamma$ to $\partial\Omega_\epsilon^+$ and $\partial\Omega_\epsilon^-$. Let $W(\textbf{x})$ be a test function. Multiplying the Eq. (\ref{SharpEq}) by the function $W(\textbf{x})$ and integrating the resulting equation over $\Omega_\epsilon$, we have
\begin{eqnarray}\label{integralEq}
    \int_{\Omega_\epsilon} \frac{\partial \psi}{\partial t}W d\textbf{x}=\int_{\Omega_\epsilon} M \nabla^2 \psi W d\textbf{x}-\int_{\Omega_\epsilon} M \nabla \cdot\int_\Gamma \textbf{n}\delta(\textbf{x}-\textbf{X})dSW d\textbf{x}.
\end{eqnarray}
The second term on the right hand side of the above equation becomes
\begin{eqnarray}
   &&\int_{\Omega_\epsilon} M \nabla \cdot\int_\Gamma \textbf{n}\delta(\textbf{x}-\textbf{X})dSW d\textbf{x} \nonumber\\
   &&= \int_{\partial \Omega_\epsilon} M \int_\Gamma \textbf{n}\delta(\textbf{x}-\textbf{X})dS W \textbf{n}_b dS_b-\int_{\Omega_\epsilon} M \int_\Gamma   \textbf{n}\delta(\textbf{x}-\textbf{X})dS \cdot \nabla W d\textbf{x}\nonumber\\
   &&=-M\int_\Gamma \textbf{n}\cdot \nabla W dS,
\end{eqnarray}
where $\textbf{n}_b$ is the outward normal vector at the domain boundary $\partial \Omega_\epsilon$. The first integral term on the right hand side of the Eq. (\ref{integralEq}) is equal to the sum of integrals of two subdomains
\begin{eqnarray}\label{twoint}
    \int_{\Omega_\epsilon} M \nabla^2 \psi W d\textbf{x}=\int_{\Omega_\epsilon^+} M \nabla^2 \psi W d\textbf{x}+\int_{\Omega_\epsilon^-} M \nabla^2 \psi W d\textbf{x}.
\end{eqnarray}
By using Green's theorem repeatedly, we have
\begin{eqnarray}
    &&\int_{\Omega_\epsilon^+} M \nabla^2 \psi W d\textbf{x}=\int_{\partial\Omega_\epsilon^+} M \nabla \psi \cdot \textbf{n}_b W dS_b+\int_\Gamma M \nabla \psi \cdot \textbf{n} W dS \nonumber \\
    &&-\int_{\partial\Omega_\epsilon^+} M \psi \nabla W \cdot \textbf{n}_b dS_b-\int_\Gamma M \psi \nabla W \cdot \textbf{n} dS+\int_{\Omega_\epsilon^+} M \nabla^2 W \psi d\textbf{x},\label{Int+}\\
    &&\int_{\Omega_\epsilon^-} M \nabla^2 \psi W d\textbf{x}=\int_{\partial\Omega_\epsilon^-} M \nabla \psi \cdot \textbf{n}_b W dS_b+\int_\Gamma M \nabla \psi \cdot (-\textbf{n}) W dS \nonumber \\
    &&-\int_{\partial\Omega_\epsilon^-} M \psi \nabla W \cdot \textbf{n}_b dS_b-\int_\Gamma M \psi \nabla W \cdot (-\textbf{n}) dS+\int_{\Omega_\epsilon^-} M \nabla^2 W \psi d\textbf{x}.\label{Int-}
\end{eqnarray}
Plugging Eqs. (\ref{Int+}) and (\ref{Int-}) into Eq. (\ref{twoint}), we get
\begin{eqnarray}
    \int_{\Omega_\epsilon} M \nabla^2 \psi W d\textbf{x}=\int_{\Gamma} M [[\nabla\psi\cdot\textbf{n}]]_\Gamma W dS-\int_{\Gamma} M [[\psi]]_\Gamma \nabla W \cdot \textbf{n} dS+\int_{\Omega_\epsilon} M \nabla^2 W \psi d\textbf{x},
\end{eqnarray}
where $[[\psi]]_\Gamma=\psi^+-\psi^-$ is the jump function across the interface $\Gamma$. As $\epsilon$ approaches zero, one can obtain
\begin{eqnarray}
    &&\int_{\Omega_\epsilon} \frac{\partial \psi}{\partial t}W d\textbf{x}\rightarrow 0, \\
    &&\int_{\Omega_\epsilon} M \nabla^2 W \psi d\textbf{x}\rightarrow 0.
\end{eqnarray}
The Eq. (\ref{integralEq}) becomes
\begin{eqnarray}\label{integralEqfinal}
    \int_\Gamma \{[[\nabla\psi\cdot\textbf{n}]]_\Gamma W + (1-[[\psi]]_\Gamma)\textbf{n}\cdot \nabla W\} dS=0.
\end{eqnarray}
Note that the fundamental lemma of calculus of variations can not be used directly to dealing with the above Eq. (\ref{integralEqfinal}). In order to achieve the goal, the test function $W$ is chosen as
\begin{eqnarray}\label{specialtest}
    W(\textbf{x})=\left\{
    \begin{aligned}
       & W_0(\textbf{x})\phi_1(\textbf{x})\exp(-\frac{1}{\phi_2(\textbf{x})})\exp(-\frac{1}{\phi_3(\textbf{x})})            &&   \textbf{x} \in \Omega_\epsilon \verb|\| \partial\Omega_\epsilon    \\
       & 0 &&\textbf{x} \in \partial\Omega_\epsilon  \\
    \end{aligned}
    \right.
\end{eqnarray}
where $W_0(\textbf{x})\in C^{\infty}(\Omega_\epsilon)$ is an arbitrary infinitely differentiable function. $\phi_1(\textbf{x})$, $\phi_2(\textbf{x})$ and $\phi_3(\textbf{x})$ are the signed distance functions. They are defined as
\begin{eqnarray}
    &&\phi_1(\textbf{x},t)=\left\{
                         \begin{array}{ll}
                           d_{min}(\textbf{x},\Gamma), &  \textbf{x} \in \Omega_\epsilon^+, \\
                           0, & \textbf{x} \in \Gamma, \\
                           -d_{min}(\textbf{x},\Gamma), & \textbf{x} \in \Omega_\epsilon^-.
                         \end{array}
                       \right.\\
    &&\phi_2(\textbf{x},t)=\left\{
                         \begin{array}{ll}
                           d_{min}(\textbf{x},\partial\Omega_\epsilon^+), &  \textbf{x} \in \Omega_\epsilon, \\
                           0, & \textbf{x} \in \partial\Omega_\epsilon^+.
                         \end{array}
                       \right.\\
    &&\phi_3(\textbf{x},t)=\left\{
                         \begin{array}{ll}
                           d_{min}(\textbf{x},\partial\Omega_\epsilon^-), &  \textbf{x} \in \Omega_\epsilon, \\
                           0, & \textbf{x} \in \partial\Omega_\epsilon^-.
                         \end{array}
                       \right.
\end{eqnarray}
Here $d_{min}(\textbf{x},\Gamma)$ is the distance function from $\textbf{x}$ to the interface $\Gamma$. It is easy to check the following relations
\begin{eqnarray}
    W(\textbf{x},t)=0, \textbf{n}\cdot \nabla W(\textbf{x},t)=W_0(\textbf{x},t)\exp(-\frac{1}{\phi_2(\textbf{x})})\exp(-\frac{1}{\phi_3(\textbf{x})})~~~~\textbf{x}\in \Gamma
\end{eqnarray}
Substituting the two above relations into Eq. (\ref{integralEqfinal}), we can get
\begin{eqnarray}
    \int_\Gamma (1-[[\psi]]_\Gamma)W_0(\textbf{x},t)\exp(-\frac{1}{\phi_2(\textbf{x})})\exp(-\frac{1}{\phi_3(\textbf{x})}) dS=0
\end{eqnarray}
Based on the fundamental lemma of calculus of variations, we must have
\begin{eqnarray}\label{interfacevalue}
   [[\psi]]_\Gamma=1.
\end{eqnarray}
Then the Eq. (\ref{integralEq}) can be simplified as
\begin{eqnarray}
    \int_\Gamma [[\nabla\psi\cdot\textbf{n}]]_\Gamma W dS=0.
\end{eqnarray}
Applying the fundamental lemma of calculus of variations again, we have
\begin{eqnarray}\label{interfaceflux}
    [[\nabla\psi\cdot\textbf{n}]]_\Gamma=0.
\end{eqnarray}
The two jump conditions has been proved.

 Here we also give the diffuse interface form of the second term of right hand side of Eq. (\ref{SharpEq}). For the three-dimensional case, we assume the surface $\Gamma$ has parametric equation $\textbf{X}=\textbf{X}(r,p)$ and the first fundamental form $ds^2=g_{rr}dr^2+2g_{rp}drdp+g_{pp}dp^2$. The surface area element is $dS=\sqrt{g}drdp$, where $g=g_{rr}g_{pp}-g_{rp}^2$ is the determinant of the metric matrix. We consider the following curvilinear coordinate system $(r,p,z)$ of the Euclidean space $R^3$
\begin{eqnarray}
   && \textbf{x}(r,p,z)=\textbf{X}(r,p)+z\textbf{n}(r,p),\\
   && \textbf{x}\rightarrow (r,p,z), \\
   &&\textbf{X}\rightarrow (r_0,p_0,0).
\end{eqnarray}
The Dirac function $\delta(\textbf{x}-\textbf{X})$ in the curvilinear coordinate system $(r,p,z)$ can be expressed as
\begin{eqnarray}
  \delta(\textbf{x}-\textbf{X})=\frac{\delta(r-r_0)\delta(p-p_0)\delta(z)}{C\sqrt{g}},
\end{eqnarray}
where the parameter $C$ is
\begin{eqnarray}
    C(r,p,z)=1-2K(r,p)z+G(r,p)z^2,
\end{eqnarray}
where $G$ is the Gauss curvature. Then $\int_\Gamma\textbf{n}(\textbf{X})\delta(\textbf{x}-\textbf{X})dS$ can be calculated as
\begin{eqnarray}
  &&\int_\Gamma\textbf{n}(\textbf{X})\delta(\textbf{x}-\textbf{X})dS=\int_\Gamma\textbf{n}(r_0,p_0)\frac{\delta(r-r_0)\delta(p-p_0)\delta(z)}{C(r,p,0)\sqrt{g}}\sqrt{g}drdp \nonumber\\
&&=\textbf{n}(r_0,p_0)\delta(z)\int_\Gamma\frac{\delta(r-r_0)\delta(p-p_0)}{C(r,p,0)}drdp=\textbf{n}(r_0,p_0)\delta(z).
\end{eqnarray}
When $\delta(z)$ is replaced by the smoothed form $\delta_\xi(z)$, the diffuse interface form is obtained.

\end{document}